\documentclass[runningheads]{svjour2}

\journalname{Journal of Biological Physics}

\usepackage{graphicx}

\usepackage{amssymb}
\usepackage{amsmath}
\usepackage{pifont}        
\usepackage{mathptmx}      
\usepackage{nicefrac}      

\newcommand{\micro}{\mbox{\Pifont{psy}\small m}} 

\begin{document}

\title{Frequency-dependent electrodeformation of giant phospholipid
  vesicles in AC electric field}

\titlerunning{Frequency-dependent electrodeformation of GUVs}

\author{Primo\v{z} Peterlin}

\institute{P. Peterlin \at
University of Ljubljana, Faculty of Medicine, Institute of Biophysics,\\
Lipi\v{c}eva 2, SI-1000 Ljubljana, Slovenia,\\
Tel.: +386-1-5437612\\
Fax.: +386-1-4315127\\
\email{primoz.peterlin@mf.uni-lj.si}
}

\date{Received: date / Accepted: date}

\maketitle

\begin{abstract}
  A model of vesicle electrodeformation is described which obtains a
  parametrized vesicle shape by minimizing the sum of the membrane
  bending energy and the energy due to the electric field.  Both the
  vesicle membrane and the aqueous media inside and outside the
  vesicle are treated as leaky dielectrics, and the vesicle itself is
  modelled as a nearly spherical shape enclosed within a thin
  membrane.  It is demonstrated (a) that the model achieves a good
  quantitative agreement with the experimentally determined
  prolate-to-oblate transition frequencies in the kHz range, and (b)
  that the model can explain a phase diagram of shapes of giant
  phospholipid vesicles with respect to two parameters: the frequency
  of the applied AC electric field and the ratio of the electrical
  conductivities of the aqueous media inside and outside the vesicle,
  explored in a recent paper (S. Aranda \emph{et al.,}
  Biophys. J. 95:L19--L21, 2008).  A possible use of the
  frequency-dependent shape transitions of phospholipid vesicles in
  conductometry of microliter samples is discussed.


\end{abstract}

\section{Introduction}
\label{sec:intro}

An increasing awareness of the effects of the electromagnetic field on
biological samples \cite{HdbkBiolEffEMFields}, as well as its use in
biotechnology \cite{Zimmermann:Electromanipulation} has instigated
numerous studies on the effects of electric field on individual
biological cells.  When exposed to an AC electric field, biological
cells exhibit several connected phenomena such as orientation,
dielectrophoresis, electrorotation and deformation
\cite{Jones:Electromech,Zimmermann:2000,Gimsa:2001g}.  Of these,
deformation has probably received the least attention, despite the
fact that it can be conveniently studied on giant unilamellar
phospholipid vesicles (GUVs) \cite{Dimova:2007,Dimova:2009}, which
serve as models of biological cells.

Extensive studies of the behaviour of biological cells in the electric
field have been conducted by Schwan since the 1950's
\cite{Schwan:1957}.  However, these early studies treated cells as
rigid objects.  A theory of lipid bilayer elasticity
\cite{helfrich:1973} was needed for the first model of vesicle
electrodeformation \cite{helfrich:1974A}.  This first model predicted
a prolate spheroidal deformation; this prediction was later confirmed
both by improved theoretical treatments
\cite{Bryant:1987,winterhalter:1988} and by experiments
\cite{Harbich:1979}, conducted in a 2~kHz AC electric field.  However,
experiments where the frequency of the field was varied
\cite{Mitov:1993,Peterlin:2000a} have demonstrated that the vesicle
shape undergoes a transition from prolate shapes at low frequencies to
oblate shapes at higher frequencies.  Independently, such behaviour
was also predicted by Hyuga and co-workers
\cite{Hyuga:1991c,Hyuga:1993}, who described a prolate-to-oblate shape
transition in certain conditions with respect to the electrical
properties of the aqueous medium inside and outside the vesicle.

\begin{figure}
  \centering\includegraphics[scale=0.3]{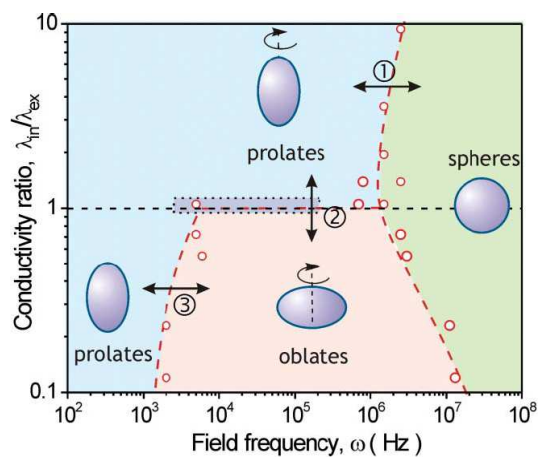}
  \caption{Morphological phase diagram of lipid vesicle shapes
    subjected to an AC electric field.  The two parameters spanning
    the diagram are the frequency of the applied AC electric field and
    the ratio of the electrical conductivities of the aqueous media
    inside and outside the vesicle (from ref.~\cite{Dimova:2007};
    reproduced by permission of The Royal Society of Chemistry).}
  \label{fig:Dimova-2007}
\end{figure}

A recent paper \cite{Aranda:2008} extended the studies of vesicle
shape in electric field to higher frequencies and explored the effect
of varying electrical conductivity of the internal and the external
aqueous medium in a systematic manner, thus obtaining a more complex
morphological phase diagram (Fig.~\ref{fig:Dimova-2007}).  The authors
have demonstrated that in the case when the conductivity of the
external aqueous medium exceeds the conductivity of the aqueous medium
inside the vesicle, another shape transitions exist in the MHz range
in addition to the previously observed prolate-to-oblate transition in
the kHz range.

The rest of this paper is structured as follows: Section
\ref{sec:theory} presents the theoretical model, which treats the
vesicle as a thin shell made of leaky dielectric and derives its shape
by minimizing its total energy, consisting of membrane bending energy
and the energy due to the electric field.  Section
\ref{sec:experiment} describes the experiments investigating the
prolate-to-oblate morphological shape transition in the kHz range.
The results -- a comparison of the model predictions with the
experimental data for the prolate-to-oblate transition on the one hand
and with the experimental morphological phase diagram
(Fig.~\ref{fig:Dimova-2007}) on the other hand -- are presented in
Section~\ref{sec:results}.  Section \ref{sec:discuss} discusses the
limitations of the presented model, compares it with the existing
ones, and introduces a possible application of frequency-dependent
transitions of vesicle shapes for conductometry.  Finally, Section
\ref{sec:conclus} presents the main conclusions.

\section{Theoretical analysis}
\label{sec:theory}

In general, the model follows the approach introduced by Winterhalter
and Helfrich \cite{winterhalter:1988}, \emph{i.e.,} a parametrized
vesicle shape is obtained by minimizing its total free energy.  Two
terms enter the free energy: the membrane bending energy and the
energy due to the electric field.  A closed-form expression for the
membrane bending energy is well-known \cite{helfrich:1973}, and the
change of the free energy due to the electric field is calculated as
the work done by the force of the electric field while deforming the
vesicle \cite{winterhalter:1988}.

Once an electric field $\mathbf{E}$ is applied, it introduces a single
distinct axis into the system, so the treatment can be limited to
axially symmetric shapes. This eliminates the dependence on the
longitude angle $\phi$, if the polar ($z$) axis is chosen parallel to
the applied field. Thus, the vesicle surface can be parametrized as
\begin{equation}
  r(\theta) = s_0 + s(\theta)\;,
\end{equation}
where $|s(\theta)| \ll s_0$, $s_0$ denoting the deformation
independent of the polar angle $\theta$.  In an absence of deformation
($|s(\theta)|=0$), $s_0$ equals the radius of the undeformed sphere
($r_0$).  The deformation is independent of the sign of the electric
field and thus proportional to $E^2$ in the lowest order.  The field
itself being proportional to $\cos\theta$, the deformation coupled
with this field can be expected to be proportional to $\cos^2\theta$.
In terms of expansion into spherical harmonics, this limits us to a
sum of even terms.  Retaining only the terms proportional to $E^2$ or
lower, a quadrupolar term remains, where $s(\theta)$ equals the second
Legendre polynomial: $s(\theta) = \frac{1}{2} s_2 (3\cos^2 \theta -
1)$, with $s_2$ being a measure for the extent of deformation.
Positive values of $s_2$ indicate prolate deformation, while negative
values indicate oblate deformation.

\subsection{Membrane bending energy}

The requirement for a local area conservation, which assures that the
membrane stretching is independent of polar angle $\theta$, implies
that a quadrupolar displacement $\delta r_r$ in a radial direction is
accompanied by a tangential displacement $\delta r_\theta$
\cite{winterhalter:1988}:
\begin{eqnarray}
  \delta r_r & = & \frac{1}{2} (3 \cos^2 \theta - 1) \,s_2
  \label{eq:displacement-r} \\
  \delta r_\theta & = & -\cos\theta
  \sin\theta \: s_2 \; .
  \label{eq:displacement-th}
\end{eqnarray}
The total membrane area expansion is determined by the relationship
between $s_0$ and $r_0$.  Taking into account the requirement for a
constant vesicle volume, the following relationship for $s_0$ is
obtained:
\begin{equation}
  s_0 = r_0 - \frac{s_2^2}{5 r_0} \; .
  \label{eq:volume-renorm}
\end{equation}
The correction for a constant volume (\ref{eq:volume-renorm}) contains
a higher term in the powers of $s_2$ and thus does not affect in the
lowest term either the bending energy or the energy due to the
electric field, yielding $s_0=r_0$ an adequate approximation.

The expression for the vesicle bending energy \cite{helfrich:1973}:
\begin{equation}
  G_\mathrm{bend} = \frac{1}{2}k_c \oint \left( c_1 + c_2 - c_0
  \right)^2 \mathrm{d}A \; .
  \label{eq.Gbend}
\end{equation}
Here, $k_c$ is the bending elastic modulus of the membrane, while
$c_1$ and $c_2$ are the principal curvatures of the membrane.  The
spontaneous curvature $c_0$ vanishes for a bilayer composed of two
equal layers.  The integration is conducted over the total membrane
area $A$ of a quadrupolarly deformed vesicle.  Up to quadratic order
terms in $s_2$, the total bending energy of a nearly spherical vesicle
can be expressed as \cite{helfrich:1973}:
\begin{equation}
  G_\mathrm{bend} = 8\pi k_c + \frac{48\pi}{5} k_c
  \left(\frac{s_2}{r_0}\right)^2 \, .
  \label{eq:Gbend}
\end{equation}
This can be readily interpreted as the bending energy of a sphere plus
an addition due to the quadrupolar deformation.

\subsection{Contribution of the electric field}

\begin{figure}
  \centering
  \includegraphics[scale=0.6]{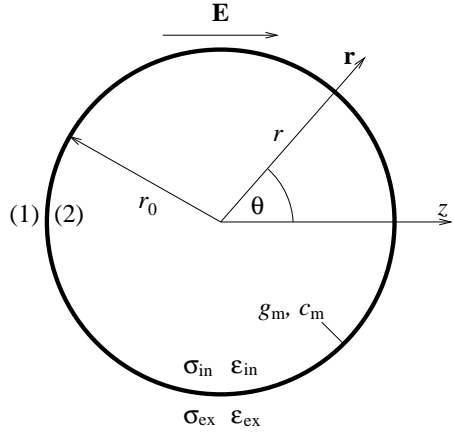}
  \caption{A vesicle is modelled in spherical coordinates $(r,\theta)$
    as a spherical shell with a radius $r$ exposed to an external
    electric field $\mathbf{E}$.  The conductivity and permittivity of
    the aqueous solution outside are denoted by $\sigma_\mathrm{ex}$
    and $\epsilon_\mathrm{ex}$, and the corresponding quantities in
    the vesicle interior by $\sigma_\mathrm{in}$ and
    $\epsilon_\mathrm{in}$. Membrane surface transconductance and
    surface capacitance are denoted by $g_m$ and $c_m$, respectively.}
  \label{fig:thin-shell}
\end{figure}

The first step in evaluating the contribution of the electric field to
the total free energy is calculating the electric field.  The present
treatment is limited to small deviations of vesicle shape from the
sphere, which greatly simplifies the treatment: instead of computing
the electric field in the presence of the actual vesicle shape, one
can compute the electric field in the presence of a spherical shell.
Both the aqueous solution inside and outside the vesicle and the
vesicle membrane are treated as lossy dielectrics.  The aqueous
solutions inside and outside the vesicle usually have identical
dielectric permittivity, while their electrical conductivities can
differ (Fig.~\ref{fig:thin-shell}).

In order to compute the forces exerted by the electric field on the
vesicle, one first has to compute the electric field around a vesicle.
Gauss' law $\nabla\cdot\mathbf{D}=0$ together with the requirement for
an irrotational electric field $\nabla\times\mathbf{E}=0$, which stems
from Faraday's law for electromagnetic induction, yields the Laplace
equation for the electric potential $U$ \cite{Landau:Electrodynamics}:
\begin{equation}
  \label{eq:laplaceU}
  \nabla^2 U = 0 \; .
\end{equation}

The boundary conditions on the membrane ($r=r_0$) require that the
total surface charge density, including both free charge and
displacement charge, must vanish.  The case where a finite potential
can be supported across the membrane is known as the series admittance
limit (\cite{Jones:Electromech}, pp.~230-232).  In the spherical
geometry, this yields a system of equations
\begin{align}
  (\sigma_\mathrm{ex} - \mathrm{i}\,\omega\epsilon_\mathrm{ex})
  E_r^{(1)}(r_0) &= (\sigma_\mathrm{in} -
  \mathrm{i}\,\omega\epsilon_\mathrm{in}) E_r^{(2)}(r_0)
  \; , \label{eq:boundarycond1} \\
  (\sigma_\mathrm{ex} - \mathrm{i}\,\omega\epsilon_\mathrm{ex})
  E_r^{(1)}(r_0) &= (g_\mathrm{m} - \mathrm{i}\,\omega c_\mathrm{m})
  \left(U^{(2)}(r_0) - U^{(1)}(r_0)\right) \; . \label{eq:boundarycond2}
\end{align}
Apart from (\ref{eq:boundarycond1}) and (\ref{eq:boundarycond2}), the
system is constrained by two additional conditions: one requiring that
the electric field far away from the vesicle is unperturbed, and the
other requiring that the electric field is finite inside the
vesicle. Membrane surface transconductance $g_\mathrm{m}$ and surface
capacitance $g_\mathrm{m}$ are related to membrane electric
conductivity and dielectric permittivity: $g_\mathrm{m} =
\sigma_\mathrm{m}/h$ and $c_\mathrm{m} = \epsilon_\mathrm{m}/h$, $h$
being the membrane thickness.

In spherical coordinates, (\ref{eq:laplaceU}) readily decouples into
the radial and the angular part, and can be solved with the usual
ansatz:
\begin{equation}
  U^{(k)} = \frac{1}{2} \left[ \left( a^{(k)} r + \frac{b^{(k)}}{r^2}
    \right) \cos\theta\, \mathrm{e}^{-\mathrm{i}\omega t} + \text{C.C.}
  \right] \; .
  \label{eq:ansatzCC}
\end{equation}
The coefficients $a^{(k)}, b^{(k)}; k = 1,2$, which can in general be
complex to allow for a phase shift, are determined from the boundary
conditions (\ref{eq:boundarycond1},\ref{eq:boundarycond2}).

The conditions for an unperturbed field far away from the vesicle and
a finite field inside the vesicle immediately yield two coefficients:
\begin{align}
  a^{(1)} &= -E_0 \; , \label{eq:a1} \\
  b^{(2)} &= 0 \; .  \label{eq:b2}
\end{align}

The two remaining coefficients are obtained by solving
(\ref{eq:boundarycond1},\ref{eq:boundarycond2}):
\begin{align}
  a^{(2)} &= -\frac{3 E_0 r}{\Delta} \bigg[ (g_\mathrm{m}
  -\mathrm{i}\omega c_\mathrm{m}) (\sigma_\mathrm{ex}
  -\mathrm{i}\omega \epsilon_\mathrm{ex}) \bigg]  \; , \label{eq:a2} \\
  b^{(1)} &= - \frac{E_0 r^3}{\Delta} \bigg[ \sigma_\mathrm{ex}
  \sigma_\mathrm{in} + g_\mathrm{m} r (\sigma_\mathrm{ex} -
  \sigma_\mathrm{in}) - \left( \epsilon_\mathrm{ex}
    \epsilon_\mathrm{in} + c_\mathrm{m} r
    (\epsilon_\mathrm{ex} - \epsilon_\mathrm{in}) \right) \omega^2 \nonumber \\
  & \qquad - \mathrm{i} \left( g_\mathrm{m} r (\epsilon_\mathrm{ex} -
    \epsilon_\mathrm{in}) + c_\mathrm{m} r (\sigma_\mathrm{ex} -
    \sigma_\mathrm{in}) + \epsilon_\mathrm{in} \sigma_\mathrm{ex} +
    \epsilon_\mathrm{ex} \sigma_\mathrm{in} \right) \omega \bigg]
  \; , \label{eq:b1} \\
  \Delta &= 2 \sigma_\mathrm{ex} \sigma_\mathrm{in} + g_\mathrm{m} r
  (2 \sigma_\mathrm{ex} + \sigma_\mathrm{in}) - \left( 2
    \epsilon_\mathrm{ex} \epsilon_\mathrm{in} + c_\mathrm{m} r
    (2 \epsilon_\mathrm{ex} + \epsilon_\mathrm{in}) \right) \omega^2 \nonumber \\
  & \qquad - \mathrm{i} \left( g_\mathrm{m} r (2 \epsilon_\mathrm{ex}
    + \epsilon_\mathrm{in}) + c_\mathrm{m} r (2 \sigma_\mathrm{ex} +
    \sigma_\mathrm{in}) + 2 (\epsilon_\mathrm{in} \sigma_\mathrm{ex} +
    \epsilon_\mathrm{ex} \sigma_\mathrm{in}) \right) \omega \;
  . \label{eq:delta}
\end{align}

The surface density of the forces exerted on the boundary of
dielectrics by the electric field can be computed as a scalar product
of the Maxwell stress tensor and a vector normal to the membrane:
\begin{equation}
  \mathbf{f} = (\underline{T}^{(1)} - \underline{T}^{(2)})
  \mathbf{e}_r \label{eq:forcedensity} \; .
\end{equation}

The force vanishes in a homogeneous medium, but can in general be
non-zero on the boundaries of media with different electrical
properties. The Maxwell stress tensor is defined as
\begin{eqnarray}
  \underline{T} & = {\bf D}\otimes{\bf E} - \frac{1}{2}({\bf
    D}\cdot{\bf E})\underline{I} \;,
\end{eqnarray}
where $\underline{I}$ denotes the identity matrix.

Unlike in the case of the bending energy term (\ref{eq:Gbend}), for
which a closed-form expression was obtained, an approach where energy
difference is computed is employed here. $\delta G_\mathrm{field}$
denotes a small change in the energy due to the electric field, when a
sphere ($s_2=0$) is perturbed by a small deformation change $\delta
s_2$. This energy difference is calculated as the work done by the
forces of the electric field during the displacement of membrane
elements $\delta\mathbf{r}$
(equations~\ref{eq:displacement-r},\ref{eq:displacement-th}),
integrated over the entire membrane area \cite{winterhalter:1988}:
\begin{equation}
\delta G_\mathrm{field} =
- \oint (\mathbf{f}\cdot\delta\mathbf{r})\,\mathrm{d}A \; .
\label{eq:Gfield-general}
\end{equation}
The integration is conducted over a sphere, which is consistent with
the limit of small deformations ($s_2 \ll r_0$).

Substituting the coefficients (\ref{eq:a1}--\ref{eq:delta}) into
(\ref{eq:Gfield-general}) yields a lengthy expression for $\delta
G_\mathrm{field}$, which can be written as a sum of two dispersion
terms:
\begin{equation}
  \delta G_\mathrm{field} = -\frac{6\pi}{5} \epsilon_\mathrm{w} E_0^2
  r_\mathrm{out}^2 s_2 \left( \xi_\infty + \frac{\xi_1}{1 + \omega^2
      \tau_1^2} + \frac{\xi_2}{1 + \omega^2 \tau_2^2} \right) \; .
\label{eq:Gfield-omega}
\end{equation}
Three dimensionless coefficients $\xi_\infty$, $\xi_1$, and $\xi_2$,
and two characteristic times $\tau_1$ and $\tau_2$ are rather lengthy
expressions involving six different material constants:
$\epsilon_\mathrm{ex}$, $\epsilon_\mathrm{in}$, $\sigma_\mathrm{ex}$,
$\sigma_\mathrm{in}$, $c_\mathrm{m}$, $g_\mathrm{m}$, and the vesicle
radius $r_0$.

\subsection{Vesicle deformation}

For small deformations, quadrupolar deformation only induces small
perturbative changes in the vesicle bending energy and the energy due
to the electric field.
\begin{eqnarray}
  G_\mathrm{bend}(s_2) &\approx G_\mathrm{bend}(s_2=0) + \frac{1}{2}
  \left.\frac{\partial^2 G_\mathrm{bend}}{\partial
  s_2^2}\right|_{s_2=0}\!\! s_2^2
  \label{eq:Gbend-perturb} \\
  G_\mathrm{field}(s_2) &\approx G_\mathrm{field}(s_2=0) +
  \left.\frac{\partial G_\mathrm{field}}{\partial
  s_2}\right|_{s_2=0}\!\! s_2
  \label{eq:Gfield-perturb}
\end{eqnarray}
Even though the total energy of the vesicle formally depends on two
parameters, $r_0$ and $s_2$, the constraint requiring a constant
vesicle volume eliminates one degree of freedom, thus yielding
(\ref{eq:Gbend-perturb}--\ref{eq:Gfield-perturb}). It is also worth
noting that the fact that neither prolate ($s_2> 0$) nor oblate shapes
($s_2 < 0$) have a bending energy lower than those of a sphere means
that the expansion for the bending energy (\ref{eq:Gbend-perturb})
contains no linear term.

Equilibrium vesicle deformation, expressed in terms of $s_2$, can then
be calculated by minimizing the total free energy over $s_2$:
\begin{equation}
  \frac{\mathrm{d}}{\mathrm{d} s_2} \left( G_\mathrm{bend} +
    G_\mathrm{field} \right) =  0 \; .
  \label{eq:equilibrium-requirement}
\end{equation}

Expressing the equilibrium vesicle deformation $s_2$ from
(\ref{eq:equilibrium-requirement}) at given conditions yields:
\begin{equation}
  s_2 = \frac{1}{16} \frac{r_0^4 \epsilon_\mathrm{w} E_0^2}{k_c}\, \left(
    \xi_\infty + \frac{\xi_1}{1 + \omega^2 \tau_1^2} + \frac{\xi_2}{1 +
      \omega^2 \tau_2^2} \right) \; .
  \label{eq:deformation-omega}
\end{equation}
As one can see, the dependence of vesicle deformation $s_2$ on the
angular frequency $\omega=2\pi\nu$ contains two dispersion terms.  It
is worth emphasizing that the only approximation used in deriving the
expression (\ref{eq:deformation-omega}) is that of a small deformation
($s_2 \ll r_0$).  Also worth noting is the fact that the only
parameter related to membrane elasticity, $k_c$, only scales $s_2$;
frequency-dependent electrodeformation of vesicles is entirely
governed by the electrical parameters of the membrane and the aqueous
medium and by the vesicle size.

\begin{figure}
  \centering\includegraphics[angle=270]{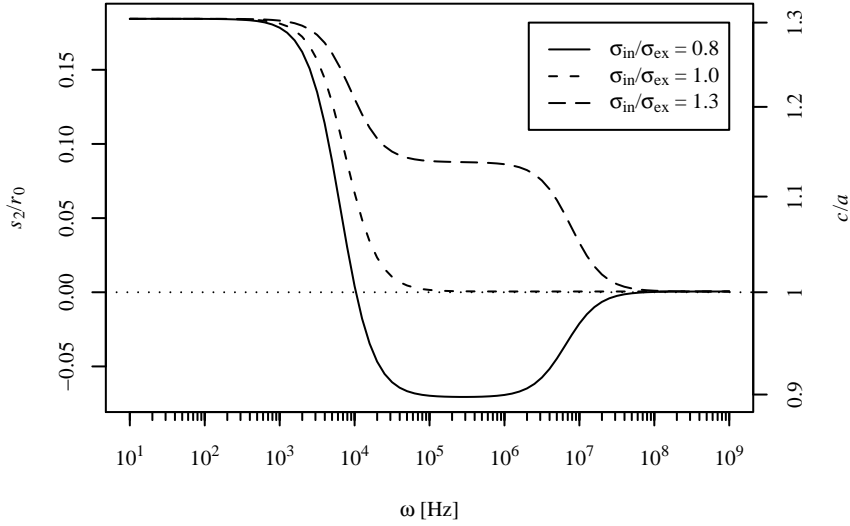}
  \caption{Vesicle deformation as a function of the frequency $\omega
    = 2\pi\nu$ of the applied AC electric field.  Vesicle deformation
    is shown both in terms of quadrupolar deformation $s_2$ normalized
    to $r_0$ (primary $y$-axis, left) and vesicle semi-axes ratio $c/a
    \approx (r_0 + s_2)/(r_0 - s_2/2)$ (secondary $y$-axis, right).
    The three curves were computed for three different ratios between
    the electrical conductivity of the external aqueous medium and the
    electrical conductivity of the internal aqueous medium.  The
    dotted line separates prolate shapes from oblate ones.  Other
    parameters used for computation were $r_0 = 20\;\micro\mathrm{m}$,
    $\sigma_\mathrm{ex} = 50\;\micro\text{S/cm}$, $\sigma_\mathrm{m} =
    10^{-14}\;\text{S/m}$, $\epsilon_\mathrm{w} = 80\, \epsilon_0$,
    $\epsilon_\mathrm{m} = 2.5\, \epsilon_0$, $h=4$~nm, $k_c = 1.2
    \times 10^{-19}$~J, $E_0 = 500$~V/m.}
\label{fig:ArandaDiagramOmega}
\end{figure}

Fig.~\ref{fig:ArandaDiagramOmega} shows vesicle deformation in
dependence of the angular frequency $\omega$ of the applied AC
electric field.  Its most prominent feature is that the deformation
exhibits a prolate-to-oblate transition at $\omega \sim
10^4\;\text{Hz}$ (\emph{i.e.,} $\nu \sim 10^3\;\text{Hz}$) in the case
when $\sigma_\mathrm{in} < \sigma_\mathrm{ex}$ (\emph{e.g.,}
$\sigma_\mathrm{in} = 0.8\,\sigma_\mathrm{ex}$, solid curve).  On the
other hand, vesicle shape remains prolate up to $\omega \sim
10^7\;\text{Hz}$ ($\nu \sim 10^6\;\text{Hz}$) when $\sigma_\mathrm{in}
> \sigma_\mathrm{ex}$ (\emph{e.g.,} $\sigma_\mathrm{in} =
1.3\,\sigma_\mathrm{ex}$, coarsely dashed curve).  In both cases, a
transition to spherical shape ($s_2 \approx 0$) is observed at $\omega
\sim 10^7\;\text{Hz}$ ($\nu \sim 10^6\;\text{Hz}$).  All these
features are in agreement with the experimental findings
\cite{Aranda:2008}.  The transition to spherical shape, however,
occurs at as low as $\omega \sim 10^4\;\text{Hz}$ when the
conductivities inside and outside are equal, $\sigma_\mathrm{in} =
\sigma_\mathrm{ex}$ (Fig.~\ref{fig:ArandaDiagramOmega}, finely dashed
curve).  Some of the points in Fig.~\ref{fig:Dimova-2007} around
$\sigma_\mathrm{in}/\sigma_\mathrm{ex} = 1$ may indicate that such
transitions have also been observed.  It needs to be pointed out,
however, that this is not a distinguished property of the
$\sigma_\mathrm{in} = \sigma_\mathrm{ex}$ case, but merely a
coincidence at given values of material properties and at given
vesicle size.

The lengthy expressions for the coefficients $\xi_\infty$, $\xi_1$,
$\xi_2$, $\tau_1$ and $\tau_2$ figuring in
(\ref{eq:deformation-omega}) can be somewhat simplified.  First of
all, it is reasonable to assume that the dielectric permittivity of
the aqueous medium inside the vesicle does not differ significantly
from the dielectric permittivity of the external medium,
$\epsilon_\mathrm{ex} = \epsilon_\mathrm{in} \equiv
\epsilon_\mathrm{w}$ (this is an approximation; it is known that the
dielectric permittivity of water decreases with the increasing
concentration of salt in it; \emph{e.g.}, the dielectric permittivity
of physiological saline is only $\approx 72\;\epsilon_0$,
\cite{Nortemann:1997}).  A further simplification is possible in the
case when realistic values for material parameters are taken into
account: (a) the conductivity of the aqueous solution greatly exceeds
that of the membrane, $\sigma_\mathrm{w} \gg \sigma_\mathrm{m}$, and
(b) with $\epsilon_\mathrm{w} \approx 80 \epsilon_0$,
$\epsilon_\mathrm{m} \approx 2.5 \epsilon_0$, and $r_0/h \sim 1000$,
then $\epsilon_\mathrm{w} \ll (r_0/h) \epsilon_\mathrm{m}$.  Retaining
only the largest terms, one obtains:
\begin{align}
  \xi_\infty &\approx \frac{\epsilon_\mathrm{w} (4 r_0 c_\mathrm{m} +
    \epsilon_\mathrm{w})} {\left( 3 r_0 c_\mathrm{m} + 2
      \epsilon_\mathrm{w} \right)^2} \; , \label{eq:xi-infty} \\
  \tau_1 &\approx \frac{\epsilon_\mathrm{w}}{\sigma_\mathrm{ex}}
  \frac{3}{2 + x} \; , \label{eq:tau1} \\
  \tau_2 &\approx \frac{r_0 c_\mathrm{m}}{\sigma_\mathrm{ex}}
  \frac{2 + x}{2 x} \; .\label{eq:tau2}
\end{align}
Here, $x = \sigma_\mathrm{in}/\sigma_\mathrm{ex}$ has been introduced
for the ratio of the conductivities of the aqueous medium inside and
outside the vesicle.

The two characteristic times are related to two distinct physical
processes: $\tau_1$ is the relaxation time for the Maxwell-Wagner
interfacial polarization, and $\tau_2$ is the charging time of a
spherical capacitor.  In the literature (see, \emph{e.g.,}
\cite{Turcu:1989a,Hyuga:1991c,Foster:1992,Jones:Electromech}) the
expressions for them are often encountered in a slightly more general
form, which distinguishes between the dielectric permittivity of the
internal and the external aqueous medium.

As expected, $\xi_\infty$ does not depend on conductivities, and the
curves for different $\sigma_\mathrm{in}/\sigma_\mathrm{ex}$ plotted
in Fig.~\ref{fig:ArandaDiagramOmega} merge into a single one at
$\omega \rightarrow \infty$.  One can also see that in the
high-frequency limit, the deformation is always slightly prolate,
$\xi_\infty \approx (h/r_0)(\epsilon_\mathrm{w}/\epsilon_\mathrm{m})
\lesssim 0.01$. Such subtle deformation is however in reality likely
to be subdued by the thermal fluctuations, and is therefore difficult
to detect experimentally.

\begin{figure}
  \centering\includegraphics[scale=0.8]{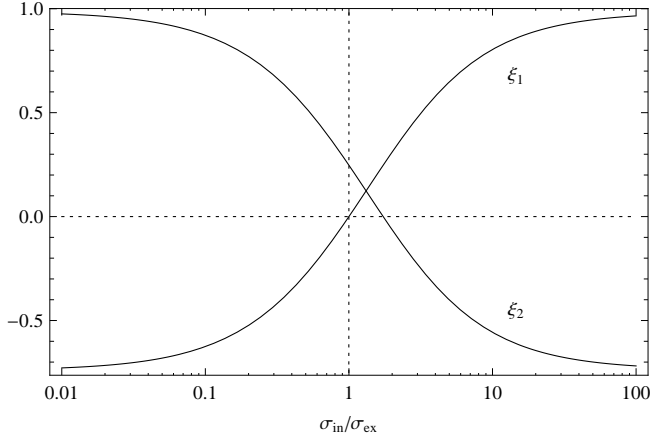}
  \caption{The coefficients $\xi_1$ and $\xi_2$ figuring in
    (\ref{eq:deformation-omega}) plotted for different values of the
    ratio of the internal and the external electrical conductivity, $x
    = \sigma_\mathrm{in}/\sigma_\mathrm{ex}$.  Other parameter values
    used for computation were the same as in
    Fig.~\ref{fig:ArandaDiagramOmega}.}
  \label{fig:ArandaDiagramXi}
\end{figure}

Fig.~\ref{fig:ArandaDiagramXi} shows how the coefficients $\xi_1$ and
$\xi_2$ change when the ratio of the internal and the external
electrical conductivity $x = \sigma_\mathrm{in}/\sigma_\mathrm{ex}$ is
varied within two decades at given values of material parameters.  On
the interval $x \in [\nicefrac{1}{100},100]$, $\xi_1$ and $\xi_2$ can
be approximated by rational functions of $x$:
\begin{align}
  \xi_1 &\approx -\frac{(r_o c_\mathrm{m} + \epsilon_\mathrm{w})(3 r_0
    c_\mathrm{m} + 4 \epsilon_\mathrm{w}) + (r_o c_\mathrm{m} +
    \epsilon_\mathrm{w})(3 r_0 c_\mathrm{m} + 4 \epsilon_\mathrm{w})
    x}{4 (r_0^2 c_\mathrm{m}^2 + 5 r_o c_\mathrm{m}
    \epsilon_\mathrm{w} + 4 \epsilon_\mathrm{w}^2) + (3 r_0^2
    c_\mathrm{m}^2 + 13 r_o c_\mathrm{m} \epsilon_\mathrm{w} + 12
    \epsilon_\mathrm{w}^2) x} \; , \label{eq:xi-1-approx} \\
  \xi_2 &\approx \frac{16 (r_0^2 c_\mathrm{m}^2 + 3 r_o c_\mathrm{m}
    \epsilon_\mathrm{w} + 2 \epsilon_\mathrm{w}^2) - (9 r_0^2
    c_\mathrm{m}^2 + 15 r_o c_\mathrm{m} \epsilon_\mathrm{w} + 4
    \epsilon_\mathrm{w}^2 ) x} {4 (3 r_0^2 c_\mathrm{m}^2 + 13 r_o
    c_\mathrm{m} \epsilon_\mathrm{w} + 4 \epsilon_\mathrm{w}^2 ) + 4
    (4 r_0^2 c_\mathrm{m}^2 + 20 r_o c_\mathrm{m} \epsilon_\mathrm{w}
    + 24 \epsilon_\mathrm{w}^2 ) x} \; . \label{eq:xi-2-approx}
\end{align}

It is possible to derive an approximate expression for the critical
value of $x$, below which the prolate-to-oblate transition occurs.  In
the intermediate region, $1/\tau_2 < \omega < 1/\tau_1$, the function
$\xi(\omega;x) = \xi_\infty + \xi_1/(1+\omega^2 \tau_1^2) +
\xi_2/(1+\omega^2 \tau_2^2)$ figuring in (\ref{eq:deformation-omega})
approximately evaluates to $\xi(\omega;x) \approx \xi_\infty + \xi_1$.
This needs to be negative in order for an oblate deformation.  From
this requirement, an approximate expression for $x_\mathrm{crit}$ can
be obtained:
\begin{equation}
  x_\mathrm{crit} \approx \frac{27\: r_0 c_\mathrm{m} + 93\:
    \epsilon_\mathrm{w}} {27\: r_0 c_\mathrm{m} + 111\:
    \epsilon_\mathrm{w}} \; .
  \label{eq:xcrit}
\end{equation}
Substituting realistic values for material parameters and vesicle size
into (\ref{eq:xcrit}), one finds that $x_\mathrm{crit} \approx 0.97$
for a vesicle with $r_0 = 5\;\micro\mathrm{m}$, $x_\mathrm{crit}
\approx 0.986$ for a vesicle with $r_0 = 10\;\micro\mathrm{m}$, and
$x_\mathrm{crit} \approx 0.997$ for a vesicle with $r_0 =
50\;\micro\mathrm{m}$.

\section{Experiment}
\label{sec:experiment}

\subsection{Vesicle preparation}

Giant unilamellar vesicles (GUVs) were prepared from commercially
available (Avanti Polar Lipids, Alabaster, AL, USA) synthetic
1-palmitoyl-2-oleoyl-\textit{sn}-gly\-cero-3-phospho\-choline (POPC)
in double-distilled water using the electroformation method
\cite{Angelova:1986,Angelova:1992} in a chamber which allows for an
easy access of the resulting vesicle suspension \cite{Peterlin:2008a}.
In order to alter the conductivity of the solution, sodium chloride
was added to the solution in concentration of up to $0.2\times
10^{-3}$ mol/L.

\subsection{Experimental procedure}
\label{subsec:exp-proc}

\begin{figure}
  \centering\includegraphics[scale=0.7]{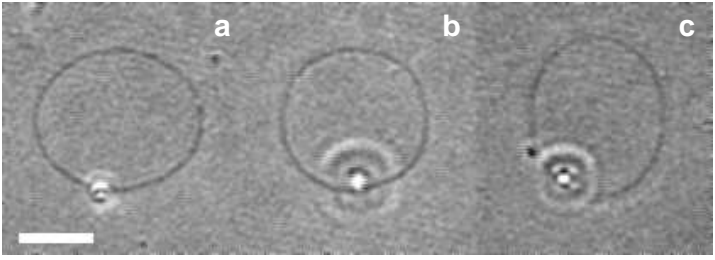}
  \caption{Phase-contrast micrograph of a vesicle in an AC electric
    field in response to the frequency of the applied field: (a)
    1~kHz, (b) 11~kHz, (c) 29~kHz.  The direction of the electric
    field corresponds to the horizontal direction.  The bar represents
    10~$\micro$m.  The conductivity of the aqueous medium is
    $\sigma_\mathrm{in} = 17\;\textrm{{\micro}S/cm}$.}
  \label{fig:collage}
\end{figure}

The experimental chamber consisted of two 0.1~mm thick stainless steel
electrodes mounted on an object glass, leaving a 0.6~mm wide gap
between the electrodes.  The cover slip was mounted with vacuum grease
(Baysilone; Bayer, Leverkusen, Germany), enclosing a few drops of
vesicle suspension.  The chamber was placed onto an inverted light
microscope (Zeiss/Opton IM 35, objective Zeiss Ph2 Plan 40/0.60;
Zeiss, Oberkochen, Germany), and phase contrast technique was used.
The micrograph was recorded using a CCD camera (Cohu 6700; Cohu, San
Diego, USA), taped using a U-Matic VCR (Sony VO-9800P) and later
digitized on a PC with a frame grabber (Matrox Meteor II; Matrox,
Dorval, Canada).  The same computer was used to set the voltage and
frequency of the applied electric field, driving a GPIB-controlled
function generator (Iskra MA-3735; Iskra, Horjul, Slovenia).  Using an
acoustic coupler connected to RS-233 port and a telephone handset with
an audio jack, the data about the selected voltage and frequency were
simultaneously recorded onto one audio channel, while the other audio
channel was left for the experimentalist's comments during the course
of experiment \cite{Sevsek:1990}.

\begin{figure}
  \centering\includegraphics[angle=270]{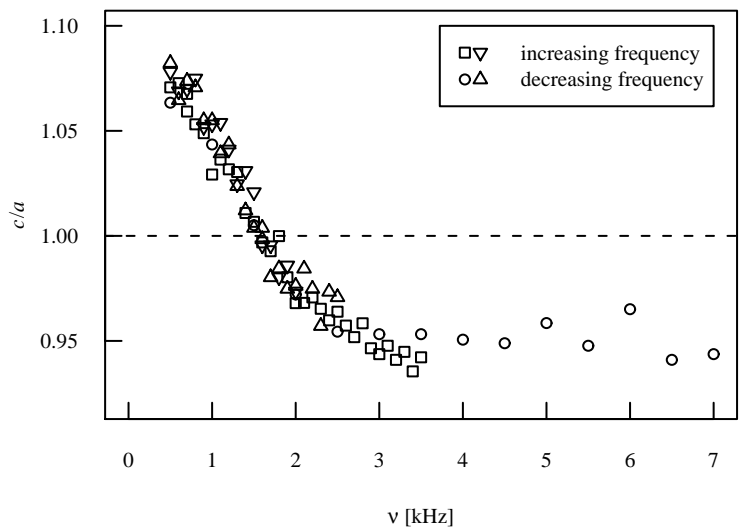}
  \caption{Prolate-to-oblate transition of a phospholipid vesicle in
    the kHz range.  Vesicle deformation is expressed as the ratio of
    its semi-axis in the direction along the field ($c$) and its
    semi-axis perpendicular to the field ($a$).  Vesicle radius $r_0 =
    19.8 \pm 0.1\;\textrm{{\micro}m}$, conductivity of the aqueous
    medium $\sigma_\mathrm{in} = 4.6\;\textrm{{\micro}S/cm}$.}
  \label{fig:vesicle}
\end{figure}

In the experiment, the prolate-to-oblate transition frequency for a
given vesicle was first coarsely estimated by testing the effect of
the electric field at 2--3 different frequencies.  After this coarse
estimate, a programmed sequence of step-wise frequency change was
applied.  The frequency step ranged from 100~Hz to 1~kHz, depending on
the vesicle, and the duration of the step was approximately 3~s.  This
duration was experimentally selected as being long enough for the
vesicle to adapt to a small frequency change, yet short enough to
minimize dielectrophoretic drift and allow repetitive measurements.
The programmed sequence was repeated several times with increasing and
decreasing frequencies in order to observe possible hysteresis
(Fig.~\ref{fig:vesicle}).  To allow a more precise determination of
transition frequency, the range was narrowed and the frequency steps
were decreased in the later runs (symbols $\square$, $\bigtriangleup$,
and $\bigtriangledown$ in Fig.~\ref{fig:vesicle}).

\section{Results}
\label{sec:results}

\subsection{Analysis of the experimental data}

\begin{figure}
  \centering\includegraphics[angle=270,scale=1.0]{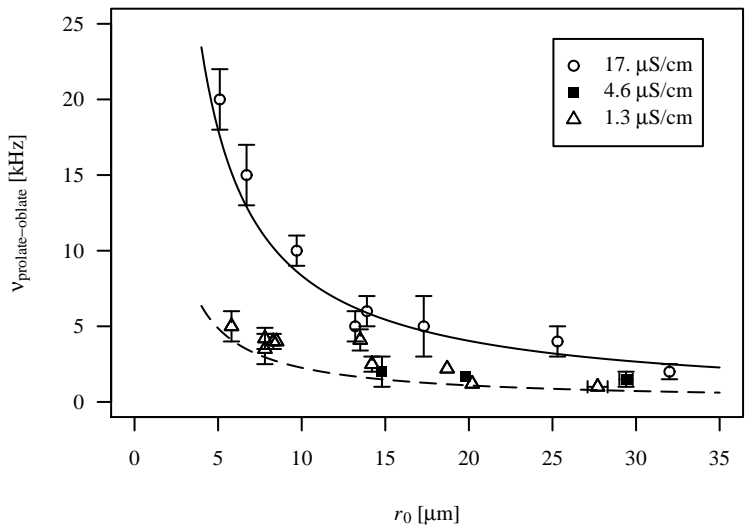}
  \caption{The dependence of the prolate-to-oblate transition
    frequency on the vesicle size and the conductivity of the aqueous
    medium.  Three runs of experiments with three different
    conductivities of the aqueous medium were performed: 17, 4.6, and
    1.3~$\micro$S/cm.  The solid line corresponds to numerically
    solving $s_2 = 0$ (Eq.~\ref{eq:deformation-omega}) for $\omega$,
    with $\sigma_\mathrm{in} = 17\;\textrm{{\micro}m}$, and the dashed
    line with $\sigma_\mathrm{in} = 4.6\;\textrm{{\micro}m}$. In both
    cases, $\sigma_\mathrm{in}/\sigma_\mathrm{ex} = 0.9$.  Other
    parameter values used for computation were the same as in
    Fig.~\ref{fig:ArandaDiagramOmega}.}
  \label{fig:ftrans-size}
\end{figure}

A total of 46 vesicle recordings were examined.  After discarding the
recordings with a vesicle relative volume too close to 1, where no
noticeable shape change was observed, as well as the recordings in
which the experiment was interrupted by the dielectrophoretic drift of
a vesicle to a region where further observations were not possible, 21
measurements on different vesicles were taken into consideration for
analysis.  Of these, 8 were prepared in a medium with conductivity
17~$\micro$S/cm, 3 in a medium with conductivity 4.6~$\micro$S/cm and
10 in a medium with conductivity 1.3~$\micro$S/cm.  The transition
frequency for the prolate-to-oblate transition vs.\ vesicle size is
plotted in Fig.~\ref{fig:ftrans-size}.  Transition frequencies are
reciprocally related to the vesicle radius, which is consistent with
the recently published findings of another group (K.\ Antonova
\emph{et al.}, in press).

The existence of prolate-to-oblate transitions indicates that the
ratio $\sigma_\mathrm{in}/\sigma_\mathrm{ex}$ in the experiments was
below $x_\mathrm{crit}$, even though the compositions of the aqueous
solution in the vesicle interior and the vesicle exterior were
initially identical.  We would like to propose an explanation for this
phenomenon.  First, we need to emphasize that the conductometer
electrode we used requires at least 6~ml of sample, therefore it was
impossible to measure directly the conductivity of the vesicle
solution.  Instead, the conductivity of the aqueous medium was
measured before it was used for hydrating the phospholipid film.  The
conductivity of the aqueous solution may increase afterwards by the
impurities introduced either at the electroformation process, during
storage or during sample preparation.  In order to test this
hypothesis, we performed a separate experiment, where we simulated the
manipulation of a vesicle sample.  Two flasks were filled with
0.03~mmol/L solution of NaCl in 0.1~mol/L sucrose solution ($\sigma =
14.5\;\textrm{{\micro}S/cm}$).  The first flask was kept sealed in
cold storage during the course of the experiment, while the second
flask was brought to room temperature every day, and a small amount of
the solution was pipetted out for conductivity measurement before
returning the flask back to cold storage.  While the conductivity of
the solution in the first flask remained unchanged, the conductivity
of the solution in the second flask increased by $\approx 8\%$ with
each consecutive pipetting, reaching $31.5\;\textrm{{\micro}S/cm}$
after 11 days.  We can presume that similar processes occur in vesicle
suspension, where the conductivity of the solution in the vesicle
interior remains unchanged as the phospholipid membrane is virtually
impermeable for ions, while the conductivity of the solution in the
vesicle exterior increases with each successive pipetting.  Since even
for the smallest of the vesicles shown in Fig.~\ref{fig:ftrans-size}
($r_0 \approx 5\;\micro\mathrm{m}$), $x_\mathrm{crit} \approx 0.97$
(\textit{cf.}  Eq.~\ref{eq:xcrit}), it is clear that at such low
conductivities of the solution, an 8\% increase of the conductivity of
the external medium caused by a single pipetting is sufficient to
bring the vesicle suspension into the regime where oblate shapes exist
in the intermediate frequency range.

The two computed lines in Fig.~\ref{fig:ftrans-size} were obtained by
numerical root-finding of $s_2 = 0$, $s_2$ being defined by
(\ref{eq:deformation-omega}).  The initial conductivity of the
solution (17~$\micro$S/cm and 4.6~$\micro$S/cm, respectively) was
taken as the conductivity of the internal aqueous medium, while the
conductivity of the external aqueous medium is
$\nicefrac{1}{0.9}\;\sigma_\mathrm{in}$, or approximately 10\% higher.
The quantitative agreement with the measured data at 17~$\micro$S/cm
and 4.6~$\micro$S/cm is within the experimental error, while the data
for 1.3~$\micro$S/cm coincide with those for 4.6~$\micro$S/cm.  This
may indicate that the actual electrical conductivity
$\sigma_\mathrm{in}$ in this latter case was not 1.3~$\micro$S/cm, but
somewhat higher (estimated around 5~$\micro$S/cm).

\subsection{Morphological diagram}

In order to plot a morphological phase diagram akin to the one shown
in Fig.~\ref{fig:Dimova-2007}, two distinct regimes have to be
considered.  For $\sigma_\mathrm{in}/\sigma_\mathrm{ex} < 1$, vesicle
deformation $s_2$ (\ref{eq:deformation-omega}) changes sign, and the
prolate-to-oblate and the oblate-to-spherical transitions can be
computed by solving $s_2 = 0$ numerically.  For
$\sigma_\mathrm{in}/\sigma_\mathrm{ex} > 1$, $s_2$ remains positive,
and instead one has to rely on the characteristic times
(\ref{eq:tau1}, \ref{eq:tau2}).  Here, the relaxation term containing
$\tau_1$ corresponds to prolate-to-spherical transition in
Fig.~\ref{fig:Dimova-2007}.  The relaxation term corresponding to
$\tau_2$ corresponds to an unobserved transition, \emph{i.e.,} a
prolate vesicle is expected to become slightly less prolate (see also
the dashed curve in Fig.~\ref{fig:ArandaDiagramOmega}) at frequencies
$\omega \gtrsim 1/\tau_2$.  Such subtle changes, however, are probably
difficult to quantify experimentally, and have therefore not been
reported in \cite{Aranda:2008}.

\begin{figure}
  \centering\includegraphics[scale=0.8]{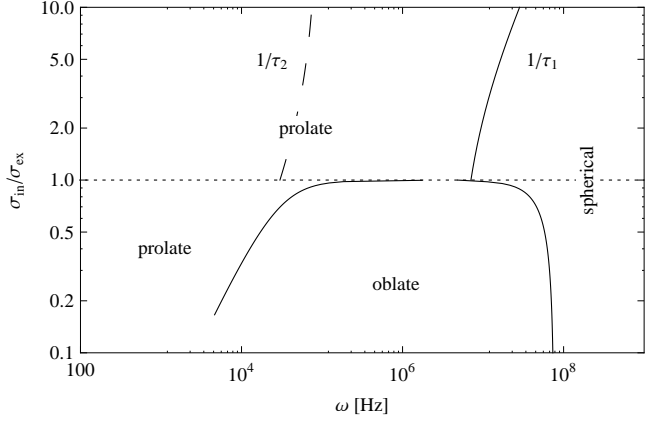}
  \caption{Prolate and oblate vesicle deformation plotted for
    different values of the ratio of the internal and the external
    electrical conductivity, $x =
    \sigma_\mathrm{in}/\sigma_\mathrm{ex}$.  In the $x < 1$ region,
    the sign changes in (\ref{eq:deformation-omega}), and either
    prolate-to-oblate transition or oblate-to-spherical transitions
    are computed, while for $x > 1$, reciprocal characteristic times
    $1/\tau_1$ and $1/\tau_2$ are plotted. Other parameter values used
    in evaluation were the same as in
    Fig.~\ref{fig:ArandaDiagramOmega}. The dashed curve corresponds to
    the relaxation which does not manifest itself as a morphological
    shape transition.}
  \label{fig:ArandaProlateOblate}
\end{figure}

The results, shown in Fig.~\ref{fig:ArandaProlateOblate}, show a
reasonable agreement with the experimental diagram
(Fig.~\ref{fig:Dimova-2007}).  In particular, for $\sigma_\mathrm{in}
< \sigma_\mathrm{ex}$, the prolate-to-oblate transition frequency
increases with increasing $\sigma_\mathrm{in}/\sigma_\mathrm{ex}$.
The prolate-to-spherical transition for $\sigma_\mathrm{in} >
\sigma_\mathrm{ex}$ also increases with increasing
$\sigma_\mathrm{in}/\sigma_\mathrm{ex}$, while the oblate-to-spherical
transition, observed for $\sigma_\mathrm{in} < \sigma_\mathrm{ex}$,
decreases with increasing $\sigma_\mathrm{in}/\sigma_\mathrm{ex}$.
All these observations are in agreement with the experiment.

We would also like to comment on the reported coexistence of prolate
and oblate vesicle shapes at $x \approx 1$ in the intermediate
frequency range \cite{Aranda:2008}.  As we have shown earlier
(Fig.~\ref{fig:ftrans-size}), the prolate-to-oblate transition
frequency exhibits a strong dependence on vesicle size: the transition
frequency for a fairly large vesicle ($r_0 = 50\;\micro\mathrm{m}$) is
$\nu_2 \approx 3\;\mathrm{kHz}$, while for a small vesicle ($r_0 =
5\;\micro\mathrm{m}$) it is $\nu_2 \approx 30\;\mathrm{kHz}$.  Within
this frequency interval, it is thus quite feasible to observe a small
prolate vesicle simultaneously with a large oblate vesicle.  While the
authors do not comment on the relative sizes of the vesicles in
question, we believe this phenomenon can be explained if: (a) the
$\sigma_\mathrm{in}/\sigma_\mathrm{ex}$ ratio was below
$x_\mathrm{crit}$ (\textit{i.e.,} slightly below 1), thus making
oblate shapes possible in the intermediate frequency range, and (b)
the frequency was above the prolate-to-oblate transition frequency for
the larger vesicle and below the prolate-to-oblate transition
frequency for the smaller vesicle.

\section{Discussion}
\label{sec:discuss}

\subsection{Comparison with earlier results}

While the model presented above builds on the legacy of the model
developed over 20 years ago by Winterhalter and Helfrich
\cite{winterhalter:1988}, it departs from it in two aspects.  Firstly,
it allows for a different electric conductivity in the vesicle
interior and the vesicle surroundings, and secondly, it is simplified
by treating the vesicle membrane as a thin shell right from the start.
The first modification was a necessity imposed by the experiments
presented in \cite{Aranda:2008}.  The limitation to thin shells is
appropriate since any results applicable to giant vesicles fall into
this limit ($\gamma \approx 1$ in \cite{winterhalter:1988}).  The
results presented here thus do not differ significantly from those
obtained by the authors who started from a general finite-thickness
model (T.\ Yamamoto \emph{et al.,} to appear).  Treating membrane
thickness as finite is, however, always required when when one
attempts to tackle membrane dielectric anisotropy
\cite{Sukhorukov:2001,Ambjornsson:2003,Ko:2004,Simeonova:2005,Peterlin:2007}.

Hyuga \emph{et al.} \cite{Hyuga:1991c} provide treatment both for the
vesicle deformation in an AC electric field and the response of the
vesicle to a step-wise change in $E$, and arrive at the expressions
(\ref{eq:tau1}, \ref{eq:tau2}).  However, they later conclude that the
fast mode dampens very quickly and that only the effects of the slow
mode are expected to show a visible influence on deformation dynamics,
and consequently focus on the lower (\emph{i.e.,} prolate-to-oblate)
transition.

More recently, a theoretical analysis of the experimental results
presented in \cite{Aranda:2008} has been published by the same group
\cite{Vlahovska:2009}, where a force-balance approach is used instead
of energy minimization.  While this does not introduce any major
advantage in the problems treated here, \emph{i.e.,} the phase diagram
of vesicle shapes, it allowed a more correct prediction of not only
the nature of deformation, but also of its extent (see below).  Even
more important, by extending the treatment beyond the equilibrium
shape it opens a possibility of treating vesicle dynamics.

Finally, we find it necessary to relate the results in this paper with
our own earlier results \cite{Peterlin:2007}, where we argued that the
prolate-to-oblate transition can be explained by the dielectric
anisotropy of the membrane.  In view of later experimental findings
\cite{Aranda:2008}, which put the prolate-to-oblate transition into a
broader context, as well as of the fact that even a very minute
difference in the electrical conductivities of the aqueous medium
inside and outside the vesicle (\emph{cf.} eq.~\ref{eq:xcrit}) makes
prolate-to-oblate transition possible, we believe it is quite possible
that the observed prolate-to-oblate transition was caused primarily by
a minute unintended increase of the conductivity of the external
aqueous medium, which may have overshadowed other possible effects.
Even though the frequency-dependent electrodeformation of giant
vesicles is likely to be primarily governed by other effects, we
however continue to believe that the dielectric anisotropy of the
phospholipid membrane is a phenomenon worth investigating.

\subsection{Quantitative estimate of vesicle deformation}

Though simple, the model presented in this paper provides a fairly
accurate description of the experimentally obtained morphological
phase diagram (Fig.~\ref{fig:Dimova-2007}), \emph{i.e.,} it predicts
the nature of deformation (prolate, oblate, or spherical), and the
frequencies at which the transitions between these shapes occur.
However, being limited by design to deformations close to spherical,
this model, like the original Winterhalter-Helfrich model, does not
predict the extent of deformation correctly.  In the case of the
Winterhalter-Helfrich model, this feature has already been noted
\cite{Aranda:2008,Vlahovska:2009}.  For the most part, the extent of
deformation is determined by the relative volume of the vesicle,
\emph{i.e.,} a flaccid vesicle deforms more than a nearly spherical
one.  For simplicity, the treatment presented in this paper omits the
effects of thermal fluctuations, which have already been treated
elsewhere \cite{Kummrow:1991,Niggemann:1995}.

\subsection{Application in microliter conductometry}

A quantitative model of vesicle shape with respect to the electrical
conductivity of the medium inside and outside the vesicle allows us to
exploit this phenomenon from the other end: a vesicle with a known
internal conductivity and a measurable size can serve as a probe for
sensing the electrical conductivity of the medium in its surroundings.
For this potential application, the kHz-range prolate-to-oblate
transition seems particularly well-suited, because (a) it is an easily
observable qualitative change in the vesicle shape which occurs in a
narrow frequency interval, and (b) the transition frequency exhibits a
strong dependence on vesicle size, which allows for some tuning,
\emph{i.e.,} in a heterogeneous vesicle population, one can select the
vesicle which manifests the effect most prominently.  Recent
developments in vesicle preparation methods
\cite{Pott:2008,Horger:2009} also bring the physiological
conductivities into reach.

\section{Conclusions}
\label{sec:conclus}

It has been demonstrated that a modified Winterhalter-Helfrich model
\cite{winterhalter:1988} which obtains a parametrized vesicle shape by
minimizing the sum of the membrane bending energy and the energy due
to the electric field, can be successfully applied to (a)
quantitatively explaining the experimentally observed
prolate-to-oblate shape transition in the kHz region, and (b) more
generally, to the experimentally obtained morphological phase diagram
of GUVs with respect to two parameters, the frequency of the applied
AC electric field, and the ratio of the electrical conductivities of
the aqueous media inside and outside the vesicle \cite{Aranda:2008}.
The obtained results are compared with the findings in the literature,
and the limitations of the model concerning the quantitative
prediction of membrane deformation are discussed.  Finally, an
application of the observed effects in the conductometry of microliter
samples is proposed.

\begin{acknowledgement}
  The author would like to thank Prof.\ S.~Svetina and Prof.\
  B.~\v{Z}ek\v{s} for numerous helpful discussions and V.~Arrigler for
  the help with vesicle preparation.  This work has been supported by
  the Slovenian Research Agency through grant J3-2268.
\end{acknowledgement}


\begin{thebibliography}{10}
\providecommand{\url}[1]{{#1}}
\providecommand{\urlprefix}{URL }
\expandafter\ifx\csname urlstyle\endcsname\relax
  \providecommand{\doi}[1]{DOI~\discretionary{}{}{}#1}\else
  \providecommand{\doi}{DOI~\discretionary{}{}{}\begingroup
  \urlstyle{rm}\Url}\fi

\bibitem{HdbkBiolEffEMFields}
Polk, C., Postow, E. (eds.): Handbook of biological effects of electromagnetic
  fields, 2nd edn.
\newblock CRC Press, Boca Raton, New York, London, Tokyo (1996)

\bibitem{Zimmermann:Electromanipulation}
Zimmermann, U., Neil, G.A. (eds.): Electromanipulation of Cells.
\newblock CRC Press, Boca Raton (1996)

\bibitem{Jones:Electromech}
Jones, T.B.: Electromechanics of particles.
\newblock Cambridge University Press, Cambridge, New York, Melbourne (1995)

\bibitem{Zimmermann:2000}
Zimmerman, U., Friedrich, U., Mussauer, H., Gessner, P., H\"amel, K.,
  Sukhorukov, V.: Electromanipulation of mammalian cells: {F}undamentals and
  application.
\newblock IEEE Trans Plasma Sci \textbf{28}, 72--82 (2000)

\bibitem{Gimsa:2001g}
Gimsa, J.: A comprehensive approach to electro-orientation, electrodeformation,
  dielectrophoresis, and electrorotation of ellipsoidal particles and
  biological cells.
\newblock Bioelectrochemistry \textbf{54}(1), 23--31 (2001)

\bibitem{Dimova:2007}
Dimova, R., Riske, K.A., Aranda, S., Bezlyepkina, N., Knorr, R.L., Lipowsky,
  R.: Giant vesicles in electric fields.
\newblock Soft Matter \textbf{3}, 817--827 (2007)

\bibitem{Dimova:2009}
Dimova, R., Bezlyepkina, N., Domange~Jord\"o, M., Knorr, R.L., Riske, K.A.,
  Staykova, M., Vlahovska, P.M., Yamamoto, T., Yang, P., Lipowsky, R.: Vesicles
  in electric fields: some novel aspects of membrane behavior.
\newblock Soft Matter \textbf{5}, 3201--3212 (2009)

\bibitem{Schwan:1957}
Schwan, H.P.: Electrical properties of tissue and cell suspensions.
\newblock Adv. Biol. Med. Phys. \textbf{5}, 147--209 (1957)

\bibitem{helfrich:1973}
Helfrich, W.: Elastic properties of lipid bilayers: Theory and possible
  experiments.
\newblock Z. Naturforsch. C \textbf{28}, 693--703 (1973)

\bibitem{helfrich:1974A}
Helfrich, W.: Deformation of lipid bilayer spheres by electric fields.
\newblock Z. Naturforsch. C \textbf{29}, 182--183 (1974)

\bibitem{Bryant:1987}
Bryant, G., Wolfe, J.: Electromechanical stresses produced in the plasma
  membranes of suspended cells by applied electric fields.
\newblock J. Membrane Biol. \textbf{96}, 129--139 (1987)

\bibitem{winterhalter:1988}
Winterhalter, M., Helfrich, W.: Deformation of spherical vesicles by electric
  field.
\newblock J. Colloid Interf. Sci. \textbf{122}, 583--586 (1988)

\bibitem{Harbich:1979}
Harbich, W., Helfrich, W.: Alignment and opening of giant lecithin vesicles by
  electric fields.
\newblock Z. Naturforsch. A \textbf{34}, 1063--1065 (1979)

\bibitem{Mitov:1993}
Mitov, M.D., M{\'e}l{\'e}ard, P., Winterhalter, M., Angelova, M.I., Bothorel,
  P.: Electric-field-dependent thermal fluctuations of giant vesicles.
\newblock Phys. Rev. E \textbf{48}(1), 628--631 (1993)

\bibitem{Peterlin:2000a}
Peterlin, P., Svetina, S., {\v Z}ek{\v s}, B.: The frequency dependence of
  phospolipid vesicle shapes in an external electic field.
\newblock Pfl{\"u}gers Arch. Eur J. Physiol. \textbf{439}, R139--R140 (2000)

\bibitem{Hyuga:1991c}
Hyuga, H., {Kinosita Jr.}, K., Wakabayashi, N.: Transient and steady-state
  deformations of a vesicle with an insulating membrane in response to
  step-function or alternating electric fields.
\newblock Jpn. J. Appl. Phys. \textbf{30}(10), 2649--2656 (1991)

\bibitem{Hyuga:1993}
Hyuga, H., {Kinosita Jr.}, K., Wakabayashi, N.: Steady-state deformation of a
  vesicle in alternating electric fields.
\newblock Bioelectrochem. Bioenerg. \textbf{32}, 15--25 (1993)

\bibitem{Aranda:2008}
Aranda, S., Riske, K.A., Lipowsky, R., Dimova, R.: Morphological transitions of
  vesicles induced by {AC} electric fields.
\newblock Biophys. J. \textbf{95}, L19--L21 (2008)

\bibitem{Landau:Electrodynamics}
Landau, L.D., Lifshitz, E.M., Pitaevski{\u\i}, L.P.: Electrodynamics of
  Continuous Media, \emph{Course of Theoretical Physics}, vol.~8, 2nd edn.
\newblock Butterworth-Heineman, Oxford (1984)

\bibitem{Nortemann:1997}
N{\"o}rtemann, K., Hilland, J., Kaatze, U.: Dielectric properties of aqueous
  {NaCl} solutions at microwave frequencies.
\newblock J. Phys. Chem. A \textbf{101}, 6864--6869 (1997)

\bibitem{Turcu:1989a}
Turcu, I., Lucaciu, C.M.: Dielectrophoresis: a spherical shell model.
\newblock J. Phys. A.: Math. Gen. \textbf{22}, 985--993 (1989)

\bibitem{Foster:1992}
Foster, K.R., Sauer, F.A., Schwan, H.P.: Electrorotation and levitation of
  cells and colloidal particles.
\newblock Biophys. J. \textbf{63}(1), 180--190 (1992)

\bibitem{Angelova:1986}
Angelova, M.I., Dimitrov, D.S.: Liposome electroformation.
\newblock Faraday Discuss. Chem. Soc. \textbf{81}, 303--311 (1986)

\bibitem{Angelova:1992}
Angelova, M.I., Sol{\'e}au, S., M{\'e}l{\'e}ard, P., Faucon, J.F., Bothorel,
  P.: Preparation of giant vesicles by external {AC} electric fields.
  {K}inetics and applications.
\newblock Prog. Colloid Polym. Sci. \textbf{89}, 127--131 (1992)

\bibitem{Peterlin:2008a}
Peterlin, P., Arrigler, V.: Electroformation in a flow chamber with solution
  exchange as a means of preparation of flaccid giant vesicles.
\newblock Colloid. Surface. B \textbf{64}, 77--87 (2008)

\bibitem{Sevsek:1990}
Sev\v{s}ek, F., Sukharev, S., Svetina, S., \v{Z}ek\v{s}, B.: The shapes of
  phospholipid vesicles in electric field as determined by video microscopy.
\newblock Stud. Biophys. \textbf{138}, 143--146 (1990)

\bibitem{Sukhorukov:2001}
Sukhorukov, V.L., Meedt, G., K\"urscher, M., Zimmerman, U.: A single-shell
  model for biological cells extended to account for the dielectric anisotropy
  of the plasma membrane.
\newblock J. Electrostat. \textbf{50}, 191--204 (2001)

\bibitem{Ambjornsson:2003}
Ambj{\"o}rnsson, T., Mukhopadhyay, G.: Dipolar response of an ellipsoidal
  particle with an anisotropic coating.
\newblock J. Phys. A: Math. Gen. \textbf{36}, 10,651--10,665 (2003)

\bibitem{Ko:2004}
Ko, Y.T.C., Huang, J.P., Yu, K.W.: The dielectric behaviour of single-shell
  spherical cells with a dielectric anisotropy in the shell.
\newblock J. Phys.: Condens. Matter \textbf{16}, 499--509 (2004)

\bibitem{Simeonova:2005}
Simeonova, M., Gimsa, J.: Dielectric anisotropy, volume potential anomalies and
  the persistent {Maxwellian} equivalent body.
\newblock J. Phys.: Condens. Matter \textbf{17}, 7817--7831 (2005)

\bibitem{Peterlin:2007}
Peterlin, P., Svetina, S., {\v Z}ek{\v s}, B.: The prolate-to-oblate shape
  transition of phospholipid vesicles in response to frequency variation of an
  {AC} electric field can be explained by the dielectric anisotropy of a
  phospholipid bilayer.
\newblock J. Phys.: Condens. Matter \textbf{19}, $136\,220$ (2007)

\bibitem{Vlahovska:2009}
Vlahovska, P.M., Serral~Graci\`a, R., Aranda-Espinoza, S., Dimova, R.:
  Electrohydrodynamic model of vesicle deformation in alternating electric
  field.
\newblock Biophys. J. \textbf{96}, 4789--4803 (2009)

\bibitem{Kummrow:1991}
Kummrow, M., Helfrich, W.: Deformation of giant lipid vesicles by electric
  fields.
\newblock Phys. Rev. A \textbf{44}(12), 8356--8360 (1991)

\bibitem{Niggemann:1995}
Niggemann, G., Kummrow, M., Helfrich, W.: The bending rigidity of
  phosphatidylcholine bilayers: Dependences on experimental method, sample cell
  sealing and temperature.
\newblock J. Phys. II France \textbf{5}, 413--425 (1995)

\bibitem{Pott:2008}
Pott, T., Bouvrais, H., M\'el\'eard, P.: Giant unilamellar vesicle formation
  under physiologically relevant conditions.
\newblock Chem. Phys. Lipids \textbf{154}, 115--119 (2008)

\bibitem{Horger:2009}
Horger, K.S., Estes, D.J., Capone, R., Mayer, M.: Films of agarose enable rapid
  formation of giant liposomes in solutions of physiologic ionic strength.
\newblock J. Am. Chem. Soc. \textbf{131}, 1810--1819 (2009)

\end{thebibliography}









\end{document}